\documentclass[12pt,a4,pdf]{article}
 \pdfoutput=1
 \usepackage{amssymb,amsmath}
\usepackage{graphicx}
\usepackage{subfig}
\usepackage{color}
\usepackage{slashed}
\definecolor{Red}{rgb}{1.,0.,0.}

\begin{document}

\begin{titlepage}
  \newcommand{\AddrLNF}{ {\it INFN, Laboratori Nazionali di
      Frascati,C.P. 13, I00044 Frascati, Italy}}
  \newcommand{\AddrJSI}{{\it J. Stefan Institute, 1000 Ljubljana, Slovenia}}
  \newcommand{\AddrHam}{{\it II. Institut f\"ur Theoretische Physik,
    Universit\"at Hamburg, Luruper Chaussee 149, 22761 Hamburg,
    Germany}}
\newcommand{\AddrLiege}{{\it Universite de Liege, 
Institut de physique Bat B5, Sart Tilman B-4000 Liege 1, Belgium}}
\vspace*{1.5cm}
\begin{center}
  \textbf{\large Implications of Flavor Dynamics for Fermion Triplet
    \vspace{0.4cm}\\
    Leptogenesis}
  \\[10mm]
  D. Aristizabal Sierra$^{a,b,}$\footnote{e-mail address: Diego.Aristizabal@lnf.infn.it},
  Jernej F. Kamenik$^{c,}$\footnote{e-mail address: jernej.kamenik@ijs.si}
  and Miha Nemev\v{s}ek$^{d,c,}$\footnote{e-mail address: miha.nemevsek@ijs.si}
  \vspace{0.8cm}
  \\
  $^a$\AddrLiege.\vspace{0.4cm}\\
  $^b$\AddrLNF.\vspace{0.4cm}\\
  $^c$\AddrJSI.\vspace{0.4cm}\\
  $^d$\AddrHam.\vspace{0.4cm}\\
\end{center}
\vspace*{0.5cm}
\begin{abstract}
  We analyze the importance of flavor effects in models in which
  leptogenesis proceeds via the decay of Majorana electroweak
  triplets. We find that depending on the relative
  strengths of gauge and Yukawa reactions the $B-L$ asymmetry can
  be sizably enhanced, exceeding in some cases an order of magnitude
  level. We also discuss the impact that such effects can have for TeV-scale
  triplets showing that as long as the $B-L$ asymmetry is produced by
  the dynamics of the lightest such triplet they are negligible,
  but open the possibility for scenarios in which the asymmetry is
  generated above the TeV scale by heavier states, possibly
  surviving the TeV triplet related washouts. We investigate these cases
  and discuss how they can be disentangled by using Majorana triplet collider
  observables and, in the case of minimal type III see-saw models even through 
  lepton flavor violation observables.
\end{abstract}
\end{titlepage}
\setcounter{footnote}{0}

\section{Introduction}
\label{sec:intro}
Measurements from light element abundances and the cosmic microwave
background radiation allow to determine the cosmic baryon asymmetry,
$Y_{\Delta_B}=(8.75\pm0.23)\times 10^{-11}$~\cite{Hinshaw:2008kr}.
The conditions under which this asymmetry can be dynamically generated
(baryogenesis) are well known~\cite{Sakharov:1967dj} and depending on
how they are realized different mechanisms for baryogenesis can be
defined~\cite{Dolgov:1991fr}.

Leptogenesis is a mechanism in which an initial lepton asymmetry is
partially reprocessed into a baryon asymmetry by nonperturbative
sphaleron interactions~\cite{Kuzmin:1985mm}. Qualitatively a net
non-zero lepton asymmetry can be generated in any framework containing
interactions that: ($i$) break lepton number; ($ii$) violate CP;
($iii$) depart from thermal equilibrium at some stage in the cosmic
evolution.  In principle, these conditions are satisfied in any
neutrino mass model in which light neutrinos acquire Majorana masses
and therefore in leptogenesis two unrelated puzzles, the origin of the
Cosmic baryon asymmetry and of neutrino masses, are linked together.
In the standard approach, in which light neutrinos acquire Majorana
masses via the type I see-saw model~\cite{seesaw}, leptogenesis takes
place via the CP violating out-of-equilibrium decays of heavy standard
model fermionic singlets~\cite{Fukugita:1986hr}. This scenario for
leptogenesis has been widely discussed in the literature
\cite{Davidson:2008bu} and indeed a lot of progress in the
understanding of the generation of the lepton asymmetry within this
scenario has been achieved~\cite{ba00,Giudice:2003jh,Abada:2006ea,
  Nardi:2006fx,Nardi:2005hs,DiBari:2005st,Engelhard:2006yg,Antusch:2010ms}.

Given that qualitatively the conditions for producing a lepton
asymmetry are also fulfilled in other see-saw realizations (type II
\cite{Schechter:1980gr} and type III~\cite{Foot:1988aq}) one is
tempted to extend the standard type I analysis to
these cases, and in fact studies of such scenarios have been
considered~\cite{Hambye:2003ka,Antusch:2004xy,Antusch:2007km,
  Hambye:2003rt,Fischler:2008xm,Strumia:2008cf}. In these models the lepton
asymmetry is generated via the dynamics of either a scalar (type II)
or a Majorana fermion (type III) $SU(2)$ triplet and thus the major difference
between these cases and the standard one arises from the fact that
both, the scalar and fermion, couple to standard model (SM) gauge
bosons. At high temperatures gauge reactions are much faster than
the expansion rate of the Universe implying that no asymmetry can be
generated at this stage. As the temperature drops, thermalization
of the triplet distribution becomes less efficient and depending on the
strength of the Yukawa interactions the generation of the lepton
asymmetry can proceed either after the decoupling of gauge reactions or after
the Yukawa interactions freeze-out~\cite{Hambye:2003rt}.

In this paper we focus on leptogenesis in type III see-saw (fermionic
triplet leptogenesis).  In particular, we analyze the dynamics of
lepton flavor in the generation of the lepton asymmetry and the
implications that such effects could have for TeV-scale triplets. As in the
standard case when the lightest triplet mass ($M_{T_1}$) is below
$10^{12}$ GeV the charged lepton Yukawa interactions that are
sufficiently fast project the lepton and anti-lepton quantum states
produced in $T_1$ decays into their flavor components before they can
re-scatter~\cite{ba00,Abada:2006ea,Nardi:2006fx}. In the standard case
the fermionic singlet dynamics is completely determined by Yukawa
reactions and accordingly the impact of flavor effects becomes well pronounced.
In contrast, in the fermionic triplet leptogenesis case, since the
dynamics of the triplet is not completely determined by its Yukawa
interactions, it is not entirely clear whether light flavor effects
could yield a sizable enhancement of the final lepton asymmetry
regardless of the strength of the Yukawa reactions. Here we will show
that the inclusion of flavor may in general produce an enhancement of
the asymmetry, but that relevant effects are possible only when the
strength of the triplet Yukawa interactions are such that they are
still active when gauge reactions decouple.

In the case of a quasi-degenerate fermionic singlet mass spectrum the
resonant enhancement of the CP violating asymmetry allows for successful
leptogenesis even when the fermionic singlet masses are well below 1
TeV~\cite{Pilaftsis:2003gt,Pilaftsis:2005rv}. In fermionic triplet
leptogenesis on the other hand, the efficiency due to gauge reactions strongly depends on
$M_{T_1}$ and it is drastically diminished when ${\cal
  O}(M_{T_1})\sim$ TeV. In the one flavor approximation in
ref.~\cite{Strumia:2008cf} it was pointed out that this constraint in
conjunction with sphaleron decoupling implies that the correct amount of
baryon asymmetry can only be generated for $M_{T_1}\gtrsim$ 1.6 TeV. As we
will show, as long as leptogenesis proceeds through $T_1$ dynamics,
this bound remains valid even when flavor effects are accounted
for. However, this does not necessarily imply that the observation of fermionic
triplets at the LHC would exclude the possibility of successful leptogenesis. 
Once flavor effects are taken into account the
possibility of high scale leptogenesis remain plausible. As will be
discussed, in that case, the $B-L$ asymmetry produced above the TeV-scale (e.g. in $T_2$ decays)
can survive $T_1$ related washouts only for particular flavor
structures. These structures constrain the dynamics of $T_1$ leading
to experimental signatures which could be used to constrain leptogenesis models. 

The rest of this paper is organized as follows. In section
\ref{sec:generalities} we discuss the generalities of type III see-saw
and review the generation of the $B-L$ asymmetry in the one flavor
approximation, we also determine the regions where Yukawa reactions
remain active after gauge interactions decouple. In section
\ref{sec:flavor-effects} we study the generation of the flavored
$B-L_i$ asymmetries showing how the inclusion of light flavors in the
analysis may impact the final asymmetry. Section
\ref{sec:implications-TeV-lepto} is devoted to the analysis of high
scale ($T_2$) leptogenesis, section \ref{sec:3.2} to the possible
collider patterns of $T_1$, induced by requiring that the asymmetry
generated at a higher scale survive $T_1$ related washouts, and in
section \ref{sec:3.3} we discuss the interplay between conditions for
successful high scale leptogenesis and lepton flavor violating
observables present in minimal type III see-saw models. In section
\ref{sec:conclusions} we summarize and present our conclusions. In
appendix \ref{sec:formulas} we present our conventions and definitions
used in our numerical calculations.

\section{Generalities}
\label{sec:generalities}
From a bottom-up approach the type III see-saw model is a simple
extension of the SM that contains $N_T$ additional fermionic $SU(2)$
triplets ($T_\alpha$) with vanishing hypercharge \footnote{Consistency with
  neutrino data~\cite{Schwetz:2008er} requires at least two
  triplets. However, apart from this constraint, the number of fermion
  triplets is arbitrary.}. In the basis in which the Majorana mass
matrix for the triplets $M_{T_\alpha}$ is real and diagonal the interactions
induced by the new states are given by the following Lagrangian
\begin{equation}
  \label{eq:lagrangian}
  {\cal L}_T = 
  i \, \text{Tr } \overline T_\alpha \slashed{D} T_\alpha
  - \lambda_{i\alpha}^* \overline \ell_i T_\alpha \widetilde H
  - \frac{1}{4} \text{Tr } T_\alpha^\dagger C M_{T_\alpha} T_\alpha
  + \mbox{h.c.}\;,
\end{equation}
where $\ell=(\nu, l)^T$ and $H=(h^+, h^0)^T$ are the lepton and
Higgs $SU(2)$ doublets ($\widetilde H=i\tau_2 H^*$). The triplets can be written as a matrix
\begin{equation}
  \label{eq:electric-triplet}
  T_\alpha = \tau^A T^A_\alpha=
  \begin{pmatrix}
    T_\alpha^0         & \sqrt{2}T_\alpha^+\\
    \sqrt{2}T_\alpha^- & -T_\alpha^0
  \end{pmatrix}\,,
\end{equation}
where $T^0 = T^3$, $T^\pm =(T^1\mp i T^2)/\sqrt{2}$ and $C$ is the charge 
conjugation operator. In this notation, the covariant
derivative is defined as $D_\mu = \partial_\mu - i g \tau^A W_\mu^A / 2$.
The Greek indices $\alpha, \beta \ldots=1, \dots, N_T$ are used to label the
different fermionic triplet generations, Latin indices $i,j, \dots$
for the lepton flavors $e, \mu, \tau$.

The Majorana triplets' mass terms break lepton number and thus,
after electroweak symmetry breaking, induce Majorana neutrino masses
for the left-handed neutrinos. In the basis
$(\nu_{Li},T^0_\alpha)$, where the heavy neutrino mass matrix
is diagonal, the neutral $(3+N_T)\times (3+N_T)$ 
fermion mass matrix reads
\begin{equation}
  \label{eq:neutral-fermion-massmatrix}
  \pmb{M_N}=
  \begin{pmatrix}
    0                  &  v\; \pmb{\lambda}\\
    v\; \pmb{\lambda}^T        &    \pmb{\hat M_T}
  \end{pmatrix}\;,
\end{equation}
where $v$ is the SM Higgs vacuum expectation value, 
$v\simeq 174$ GeV. Accordingly the effective light neutrino mass
matrix is given by
\begin{equation}
  \label{eq:light-nmm}
  \pmb{m_\nu}^{\text eff} = -v^2 \pmb{\lambda}\;\cdot \pmb{\hat M_T}^{-1} \;\cdot
  \pmb{\lambda}^T\;.
\end{equation}
In general the Yukawa coupling matrix $\pmb{\lambda}$ is a complex
$3\times N_T$ matrix in flavor space and therefore contains new sources
of CP violation, as can be clearly seen by expressing $\pmb{\lambda}$
by means of the Casas-Ibarra parametrization~\cite{Casas:2001sr}:
\begin{equation}
  \label{eq:casas-ibarra}
  \pmb{\lambda} = \pmb{U}^*\cdot \sqrt{\pmb{\hat{m}_\nu}}\cdot\pmb{R}
  \cdot\sqrt{\pmb{\hat{M}_T}}\;,
\end{equation}
which ensures that the correct low energy parameters are obtained.
Here $\pmb{U}$ is the leptonic mixing matrix which diagonalizes the
effective neutrino mass matrix $\pmb{m_\nu}^{\text eff}$ and is
fixed by low energy observables (three light neutrino mixing
angles, one Dirac and one or two Majorana CP violating phases). The matrix
$\pmb{R}$ is an orthogonal complex matrix defined by $N_T(N_T-1)/2$ complex parameters.

The CP violating sources contained in $\pmb{\lambda}$ induce
two-body CP violating decays of $T_\alpha$. The tree level decay width
for these processes reads
\begin{equation}
  \label{eq:decay-width}
  \Gamma_\alpha
  =\sum_j\Gamma(T_\alpha\to \overline \ell_j\widetilde H,\ell_j\widetilde H^\dagger)
  =\frac{M_{T_\alpha}}{8\pi} \sum_{i=e,\mu,\tau}\lambda^*_{i\alpha}\lambda_{i\alpha}
  =\frac{M_{T_\alpha}^2}{8\pi\,v^2}\sum_{i=e,\mu,\tau}\widetilde m_{i\alpha}\;,
\end{equation}
where
\begin{equation}
  \label{eq:mtilde-def}
  \widetilde m_{i\alpha} = \frac{v^2}{M_{T_\alpha}}\lambda_{i\alpha}^*\lambda_{i\alpha}
  \quad \mbox{and} \quad
  \widetilde m_\alpha=\sum_{i=e,\mu,\tau}\widetilde m_{i\alpha}\;.
\end{equation}
In terms of (\ref{eq:casas-ibarra}) these parameters are determined by
low energy observables and by the entries of the matrix
$\pmb{R}$. Given that such entries can be arbitrarily large, without
affecting the light neutrino masses, light triplets (${\cal O}(\mbox{TeV})$)
not necessarily imply small values of $\widetilde m_\alpha$.

\begin{figure}
  \centering
  \includegraphics[width=11cm,height=2.4cm]{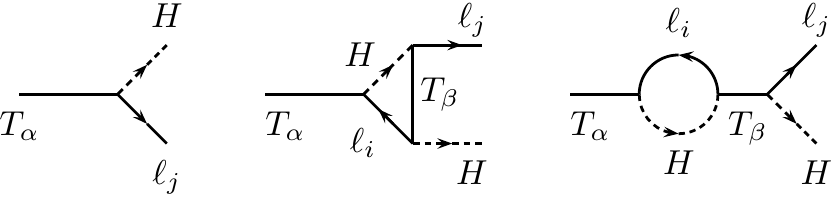}
  \caption{Diagrams generating the flavored CP violating asymmetries
    $\epsilon_{T_\alpha}^{\ell_i}$.}
  \label{fig:cp-asymm}
\end{figure}
The CP violating asymmetries for $T_\alpha$ decays into $\ell_j$
lepton flavor arise, as in type I see-saw~\cite{co96}, through the
interference between the tree level and the one-loop vertex (V) and
wave-function (W) diagrams (see fig. \ref{fig:cp-asymm}). The
corresponding expressions are given by
\begin{align}
  \label{eq:CPV-asymmetries}
  \epsilon_{T_\alpha}^{\ell_j\text{(V)}}&=\frac{1}{8\pi}\sum_{\beta\neq\alpha}
  \frac{\mathbb{I}\mbox{m}[\sqrt{\omega_\beta}
    (\lambda^\dagger\lambda)_{\beta\alpha}
    \lambda_{j\beta}^*\lambda_{j\alpha}]}
  {(\lambda^\dagger\lambda)_{\alpha\alpha}}f(\omega_\beta)
  \;,\nonumber\\
  \epsilon_{T_\alpha}^{\ell_j\text{(W)}}&=
  -\frac{1}{8\pi}\sum_{\beta\neq\alpha}
  \frac{
    \mathbb{I}\mbox{m}
    \{
    [(\lambda^\dagger\lambda)_{\alpha\beta}
    +
    \sqrt{\omega_\beta}(\lambda^\dagger\lambda)_{\beta\alpha}]
    \lambda_{j\beta}^*\lambda_{j\alpha}
    \}}
  {(\lambda^\dagger\lambda)_{\alpha\alpha}}g(\omega_\beta)\;,
\end{align}
where $\omega_\beta=M_{T_\beta}^2/M_{T_\alpha}^2$ and the loop functions
are given by
\begin{align}
  \label{eq:loop-functions}
  f(\omega_\beta)&=(1+\omega_\beta)\ln
  \left(
    \frac{\omega_\beta + 1}{\omega_\beta}
  \right) - 1\;,\nonumber\\
  g(\omega_\beta)&=\frac{\omega_\beta-1}{(\omega_\beta-1)^2+a_{\Gamma_\beta}^2}\;,
\end{align}
with $a_{\Gamma_\beta}=\Gamma_\beta/M_\alpha$. In the case of a
quasi-degenerate triplet spectrum ($\sqrt{\omega_\beta}\sim 1 +
a_{\Gamma_\beta}$) the wave-function piece is resonantly enhanced. In
such cases, depending on the size of the CP violating phases, the
flavored CP asymmetries, entirely determined by the wave-function
piece, can be ${\cal O}(1)$ (see
ref.~\cite{Pilaftsis:2003gt,Pilaftsis:2005rv} for a thorough
discussion of such possibility within type I see-saw).  In the case of
a hierarchical triplet spectrum ($\omega_\beta\gg$ 1 with
$\beta>\alpha$) $a_{\Gamma_\beta}$ can be neglected and the loop
functions expanded in powers of $\omega_\beta^{-1}$. At leading order
the flavored CP violating asymmetries can be expressed as
\begin{equation}
  \label{eq:flavored-cpv-asymmetry-leading}
  \epsilon_{T_\alpha}^{\ell_j}=
  -\frac{1}{8\pi(\lambda^\dagger\lambda)_{\alpha\alpha}}
  \sum_{\beta\neq\alpha}
  \mathbb{I}\mbox{m}
  \left\{
    \left[
      \frac{(\lambda^\dagger\lambda)_{\alpha\beta}}{\omega_\beta}
      + 
      \frac{(\lambda^\dagger\lambda)_{\beta\alpha}}{2\sqrt{\omega_\beta}}
    \right]
    \lambda_{j\beta}^*\lambda_{j\alpha}
    \right\}\,.
\end{equation}
Some comments are in order regarding the different terms in this
expression. Since the term
$\sum_\beta(\lambda^\dagger\lambda)_{\alpha\beta}
(\lambda^\dagger\lambda)_{\beta\alpha}$ is real, the first term only
contributes to the flavored CP violating asymmetries
\cite{Nardi:2006fx,co96}.  Although, apart from the kinetic term, the
Lagrangian in eq.~(\ref{eq:lagrangian}) has the same structure as
the Lagrangian in type I see-saw, the contractions of the $SU(2)$
indices in the Yukawa interaction terms are different and consequently
the second term in (\ref{eq:flavored-cpv-asymmetry-leading}) is a
factor 3 smaller than the corresponding one in the standard
case\footnote{In contrast to the type I see-saw case, due to these
  contractions, there is a relative sign between the wave-function and
  vertex pieces. Thus, at leading order,
  $\epsilon_{T_\alpha}^{\ell_j\text{(W)}}\to 1/\sqrt{\omega_\beta}$
  whereas $\epsilon_{T_\alpha}^{\ell_j\text{(V)}}\to
  -1/2\sqrt{\omega_\beta}$}.

In the strongly hierarchical limit, where the first term in
(\ref{eq:flavored-cpv-asymmetry-leading}) can be neglected, and
assuming a normal hierarchical light neutrino spectrum it can be shown
that as in the standard case there is an upper bound on the flavored
CP violating asymmetries, namely~\cite{Abada:2006ea}
\begin{equation}
  \label{eq:bound}
  \epsilon_{T_\alpha}^{\ell_j}\lesssim 10^{-5}\;
  \left(\frac{M_{T_\alpha}}{10^{10}\;\mbox{GeV}}\right)
  \left(\frac{m_3}{1\;\mbox{eV}}\right)
  \frac{\widetilde{m}_{j\alpha}}{\widetilde{m}_\alpha}\,.
\end{equation}
Finally, the total CP violating asymmetry can be obtained from
(\ref{eq:flavored-cpv-asymmetry-leading}) by summing over the flavor
index $j$~\cite{Hambye:2003rt}
\begin{equation}
  \label{eq:eps_tot-hierarchical-case}
  \epsilon_{T_\alpha}=-\frac{1}{16\pi}\sum_\beta\frac{1}{\sqrt{\omega_\beta}}
  \frac{\mathbb{I}\mbox{m}[(\lambda^\dagger\lambda)_{\beta\alpha}^2]}
  {(\lambda^\dagger\lambda)_{\alpha\alpha}}\;.
\end{equation}

\begin{figure}
  \centering
  \includegraphics[width=10cm,height=6.5cm]{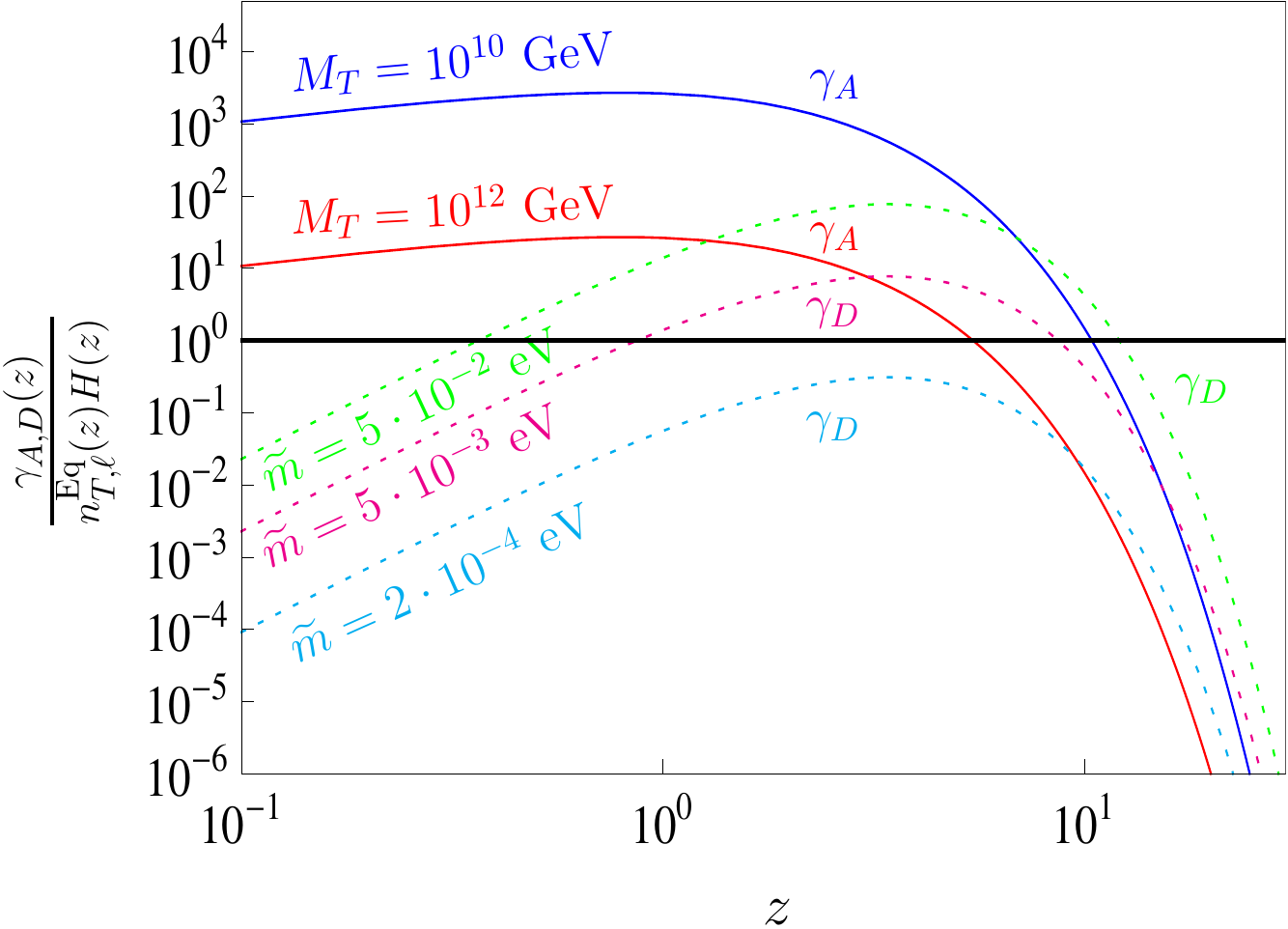}
  \caption{Evolution of gauge and Yukawa reaction densities
    ($\gamma_A$, $\gamma_D$) with $z$.  The horizontal black line
    indicates when do either gauge or Yukawa reactions go out of
    equilibrium.}
  \label{fig:guage-yukawa-reaction-dens}
\end{figure}

The major quantitative difference between the standard leptogenesis
scenario and fermion triplet leptogenesis lies in the fact that
fermion triplets couple to SM gauge bosons. At $z=M_T/T \ll 1$ gauge
reactions thermalize the triplet distribution meaning that no
asymmetry is produced at high $T$.  Gauge reactions decouple at temperatures when
\begin{equation}
  \label{eq:decoupling-condition}
  \frac{\Gamma_A}{H}=\frac{\gamma_A}{n_T^{\text{Eq}}H}\lesssim 1\,,
\end{equation}
where $\gamma_A$ is the gauge reaction density, $H$ is the expansion
rate of the Universe and $n_T^{\text{Eq}}$ is the equilibrium triplet
number density (see appendix \ref{sec:formulas}). Thus, if at this
stage inverse decay processes $\ell H \to T$ are decoupled as well
($\gamma_D/n^{\text{Eq}}_\ell H\lesssim 1$, where $\gamma_D$ is the
decay reaction density), the CP violating out-of-equilibrium decays of
the triplets will produce a $B-L$ asymmetry. Conversely, if inverse
decays are still active when gauge reactions decouple the $B-L$
asymmetry will be generated at lower temperatures, after inverse
decays switch off.  Since $\gamma_A/\gamma_D\sim g^4/M_T\widetilde m$,
as $M_T$ decreases only large values of $\widetilde m$ are able to
mantain inverse decays active after gauge reaction decoupling takes
place. Fig. \ref{fig:guage-yukawa-reaction-dens} shows an illustrative
example where it can be seen that for $M_T=10^{12}$ GeV, inverse decay
processes are still active after gauge interaction decoupling provided
$\widetilde m\sim 5\times 10^{-3}$ eV, whereas for $M_T=10^{10}$ GeV
this occurs for a $\widetilde m$ an order of magnitude
larger. Independently of how leptogenesis proceeds, a precise
determination of the generated $B-L$ asymmetry requires a numerical
treatment of the corresponding Boltzmann equations, which we now
discuss in turn.

\subsection{Generation of the $B-L$ asymmetry}
\begin{figure}
  \centering
  \includegraphics[height=6cm,width=7.7cm]{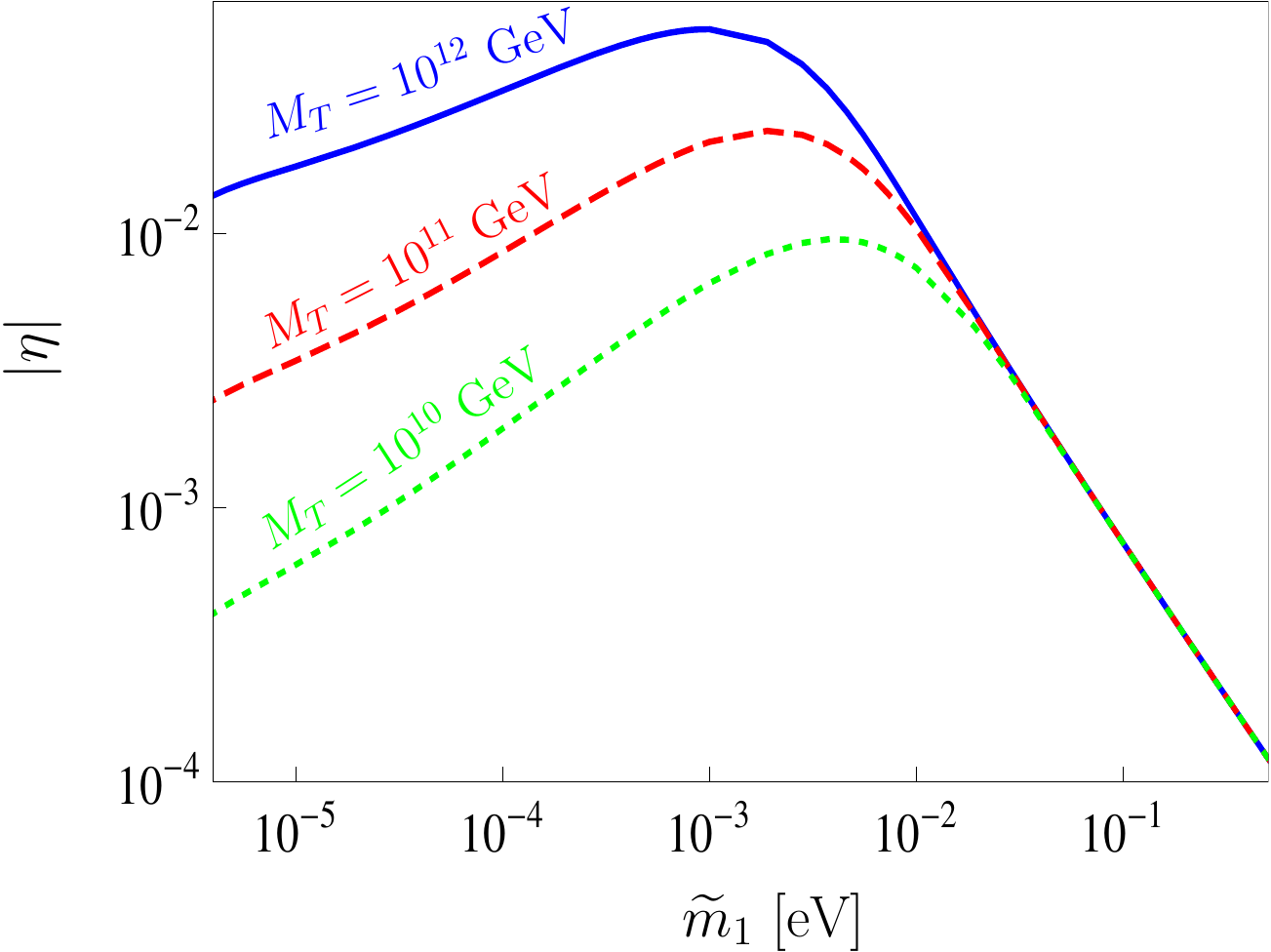}
  \includegraphics[height=6cm,width=7.7cm]{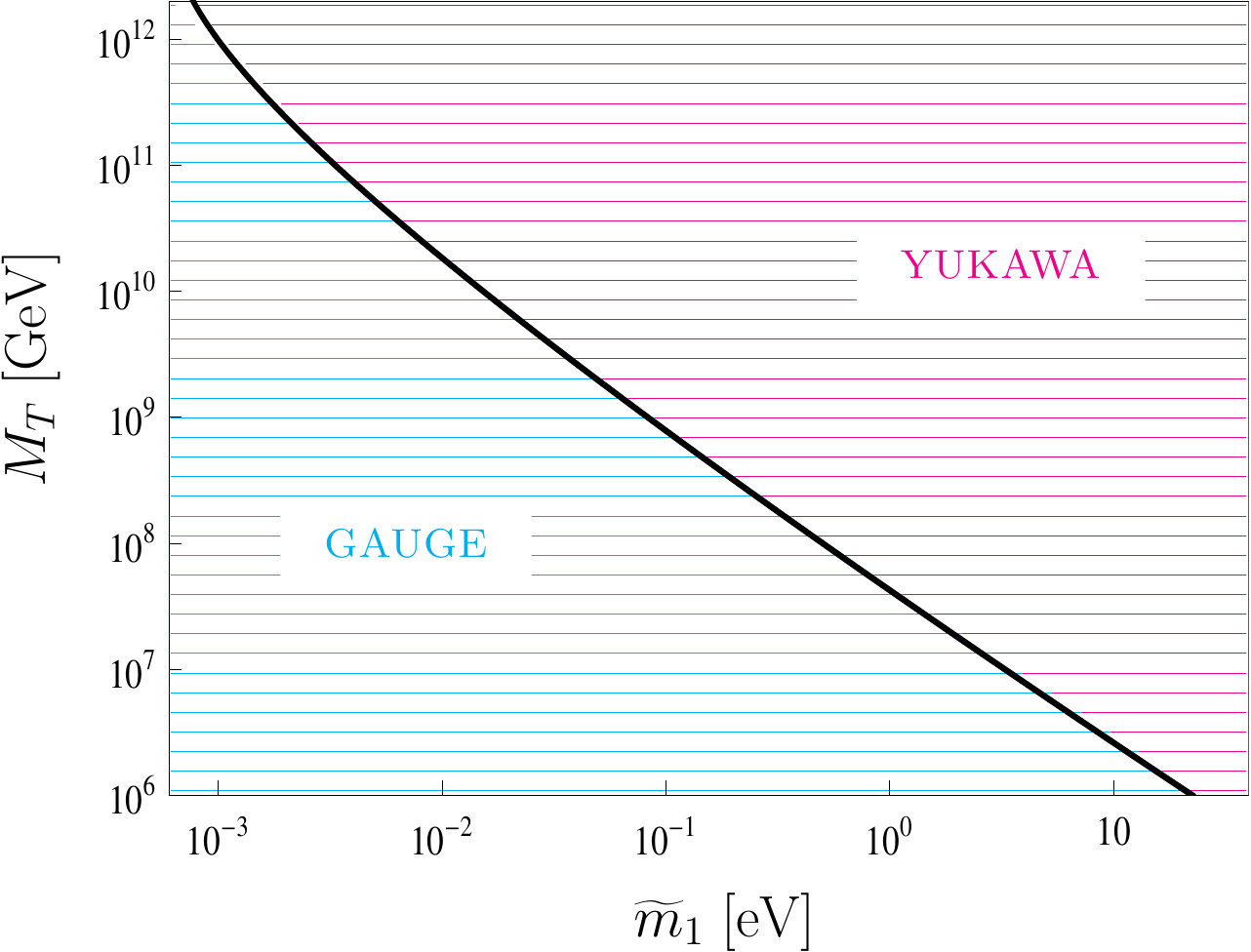}
  \caption{Efficiency factor as a function of $\widetilde m_1$ in the
    flavor aligned case (left panel) and regions for which gauge
    interactions freeze out after (lower region) and before (upper
    region) Yukawa reaction decoupling (right panel).}
  \label{fig:eff-mtilde-aligned}
\end{figure}

Although in this section we will focus on the case in which the
asymmetry is generated along one specific flavor direction, we will
write flavor dependent Boltzmann equations as they will be used in
sec. \ref{sec:flavor-effects}. Considering only triplet annihilations
(mediated by SM gauge bosons), decays (induced by the
Yukawa couplings $\pmb{\lambda}$) and $\Delta L=2$ scatterings
($\ell_j\widetilde H^\dagger\leftrightarrow \ell_i\widetilde
H^\dagger$ and $\overline\ell_j\widetilde H\leftrightarrow
\overline\ell_i\widetilde H$) \footnote{Although we work at ${\cal
    O}(\lambda^2)$ the inclusion of these processes is mandatory to
  obtain Boltzmann equations with the correct thermodynamical behavior~\cite{Nardi:2007jp}.}  the network of flavor dependent Boltzmann
equations at ${\cal O}(\lambda^2)$, assuming a vanishing Higgs number
asymmetry~\cite{Nardi:2005hs}, can be written as
\begin{align}
  \label{eq:BEQs1}
  \frac{dY_{T_\alpha}}{dz_\alpha}&=-\frac{1}{sHz_\alpha}
  \left[
    \left(
      \frac{Y_{T_\alpha}}{Y_{T_\alpha}^{\text{Eq}}}-1
    \right)\gamma_{D_\alpha}
    +
    \left(
      \frac{Y_{T_\alpha}^2}
      {\left(Y_{T_\alpha}^{\text{Eq}}\right)^2}-1
    \right)\gamma_{A_\alpha}
  \right]\,,\\
  \label{eq:BEQs2}
  \frac{dY_{\Delta_i}}{dz_\alpha}&=-\frac{1}{sHz_\alpha}
  \left[
    \left(
      \frac{Y_{T_\alpha}}{Y_{T_\alpha}^{\text{Eq}}}-1
    \right)\epsilon_{T_\alpha}^{\ell_i}
    +
    \frac{K_{i\alpha}}{2Y_\ell^{\text{Eq}}}
    \sum_{j=e,\mu,\tau}C_{ij}^\ell Y_{\Delta_j}
  \right]\gamma_{D_{\alpha}}\;.
\end{align}
Here $z_\alpha=M_{T_\alpha}/T$, $Y_X=n_X/s$ and
$Y_{\Delta_i}=Y_{\Delta_{B/3 - L_i}}$ with $Y_{L_i}=2Y_{\ell_i} +
Y_{e_i}$ (the lepton asymmetry distributed in left and right handed
degrees of freedom). The flavor projectors $K_{i\alpha}$ are defined
as follows~\cite{Nardi:2005hs}:
\begin{equation}
  \label{eq:projectors}
  K_{i\alpha}=\frac{\lambda_{i\alpha}^*\lambda_{i\alpha}}
  {(\lambda^\dagger\lambda)_{\alpha\alpha}}=
  \frac{\widetilde m_{i\alpha}}{\widetilde m_\alpha}\;,
\end{equation}
note that $\sum_{i=e,\mu,\tau} K_{i\alpha}=1$. The numerical
coefficients $C_{ij}^\ell$, that relate $Y_{\ell_i}$ with
$Y_{\Delta_i}$ according to~\cite{Nardi:2006fx}
\begin{equation}
  \label{eq:yelli-ydeltai}
  Y_{\ell_i}=-\sum_j C^\ell_{ij} Y_{\Delta_j}\;,
\end{equation}
couple the different differential equations in (\ref{eq:BEQs2}) and are
determined by the reactions that at a certain temperature regime are in
equilibrium~\cite{ba00,Nardi:2006fx,Abada:2006ea}, finally
\begin{equation}
  \label{eq:gamma-flavor}
  \gamma_{D_\alpha}=\sum_{i=e,\mu,\tau}
  \gamma_{D_{i\alpha}}\;.
\end{equation}
By integrating eq. (\ref{eq:BEQs2}) the resulting $Y_{\Delta_{B-L}}$
asymmetry can be expressed according to (see appendix
\ref{sec:formulas})
\begin{equation}
  \label{eq:BmLAsymm}
  Y_{\Delta_{B-L}}=\sum_{i=e,\mu,\tau}Y_{\Delta_i}=
  3\times\sum_{i=e, \mu, \tau}
  \epsilon_{T_{\alpha}}^{\ell_i}Y^{\text{Eq}}_{T_\alpha}\eta_{i\alpha}\;,
\end{equation}
where the factor 3 accounts for the three $SU(2)$ degrees of freedom
of the triplets and $\eta_{i\alpha}$ denotes the efficiency factor
that accounts for the $T_\alpha$ generated asymmetry in flavor $i$
that survives washouts. In the case of a hierarchical spectrum
($M_{T_1}\ll M_{T_{\alpha>1}}$), neglecting the dynamics of the heavier
triplets, leptogenesis proceeds entirely through $T_1$
dynamics. Compared with the standard case the amount of
$Y_{\Delta_{B-L}}$ is smaller and its value can strongly depend on
$M_{T_1}$ (depending on the size of $\widetilde m_{1}$). In order to
demonstrate this we have numerically solved eqs. (\ref{eq:BEQs1}) and
(\ref{eq:BEQs2}) assuming alignment along flavor $j$
($K_{j1}=\delta_{j1}$) and with $C_{ij}^\ell= \delta_{ij}$. Figure
\ref{fig:eff-mtilde-aligned} (left panel) shows the evolution of the
efficiency factor ($\eta$) with $\widetilde m_1$ for several values of
$M_{T_1}$. It can be seen that the dependence on $M_{T_1}$ is stronger
for small values of $\widetilde m_1$ and less pronounced as
$\widetilde m_1$ increases, being completely independent of $M_{T_1}$
for large values of $\widetilde m_1$, which means there is a
$\widetilde m_1^{\text{min}}$ above which leptogenesis proceeds as in
the standard case \footnote{In standard leptogenesis at ${\cal
    O}(\lambda^2)$ the efficiency does not depend on the fermionic
  singlet mass at all.}. This implies that although gauge processes
introduce a dependence on the triplet mass this dependence disappears
as soon as the gauge interactions are subdominant with respect to
Yukawa reactions. Thus, in the fermionic triplet leptogenesis the
generation of the $B-L$ asymmetry can proceed either in a region
determined by the condition $\widetilde m_1<\widetilde
m_1^{\text{min}}$ or conversely in a region defined by $\widetilde
m_1>\widetilde m_1^{\text{min}}$. The determination of $\widetilde
m_1^{\text{min}}$, for a given triplet mass, can be done as
follows. From the gauge decoupling condition $\gamma_A/n^{\text{Eq}}_T
H\lesssim 1$ the corresponding $z=z_{A-\text{dec}}$ at which gauge
reactions go out of equilibrium can be calculated. The value
$\widetilde m_1^{\text{min}}$ can be computed by requiring that the
Yukawa interactions are still active at $z_{A-\text{dec}}$ i.e
$(\gamma_D/n^{\text{Eq}}_\ell H)|_{z={z_{A-\text{dec}}}}\gtrsim 1$.
Figure \ref{fig:eff-mtilde-aligned} shows the values ($\widetilde
m_1$,$M_{T_1}$) for which the $B-L$ asymmetry either depends on
$M_{T_1}$ (``gauge region'') or is entirely determined by $\widetilde
m_{1}$ (``Yukawa region'').

\section{Including Flavor}
\label{sec:flavor-effects}
Flavor effects become relevant at temperatures below $10^{13}$ GeV
when bottom and tau Yukawa interactions enter into thermodynamic
equilibrium~\cite{ba00,Abada:2006ea,Nardi:2006fx}.  For definiteness,
from now on, we will focus in the temperature window $10^9
\;\mbox{GeV}\lesssim M_{T_1}\lesssim 10^{12} \;\mbox{GeV}$ at which, in
addition to the bottom and tau Yukawa processes, also electroweak sphalerons
are already in thermal equilibrium. 
In this regime the $B-L$ asymmetry is distributed
along $\ell_\tau$ and $\ell_1$ (an admixture of muon and electron
flavors). The determination of the total asymmetry in this case is
therefore a two flavor problem ($\ell_\tau$, $\ell_1$) and the network
of Boltzmann equations in (\ref{eq:BEQs1})-(\ref{eq:BEQs2}) consist of
three coupled differential equations in the variables $Y_{T_1}$,
$Y_{\Delta_\tau}$ and $Y_{\Delta_1}$ where the flavored asymmetries
are coupled by the flavor coupling matrix~\cite{Nardi:2006fx}
\begin{equation}
  \label{eq:coupling-matrix-T1}
  C^\ell=\frac{1}{460}
  \begin{pmatrix}
    196 & -24\\
    -9  & 156
  \end{pmatrix}\,,
\end{equation}
obtained from the chemical equilibrium conditions imposed by reactions
that in the corresponding temperature window are faster than the
expansion rate~\cite{Harvey:1990qw}. According to the discussion of the previous section, in
triplet fermion leptogenesis the asymmetry can proceed in one of the
two following {\it regimes}: $(a)$ Gauge reaction decoupling takes
place after Yukawa induced inverse decays have gone out of equilibrium --
``gauge region''; $(b)$ Gauge reactions freeze-out when inverse decay
processes are still active -- ``Yukawa region''. In scenario $(a)$ since
the the triplet abundance is efficiently diminished by gauge boson
mediated annihilations the effects of flavor are expected to be small,
or even negligible.  Instead, in scenario $(b)$ since at $z_1\gg 1$
the dynamics of $T_1$ is entirely determined by Yukawa reactions,
flavor effects may be sizable and can be prominent as long as
\begin{equation}
  \label{eq:flavor-enhancement-cond}
  \left|\epsilon_{T_1}^{\ell_j}\right|>\left|\epsilon_{T_1}^{\ell_i}\right|
  \quad\mbox{and}\quad
  K_{j1}\ll K_{i1}\,.
\end{equation}
\begin{figure}
  \centering
  \includegraphics[width=7.8cm,height=6cm]{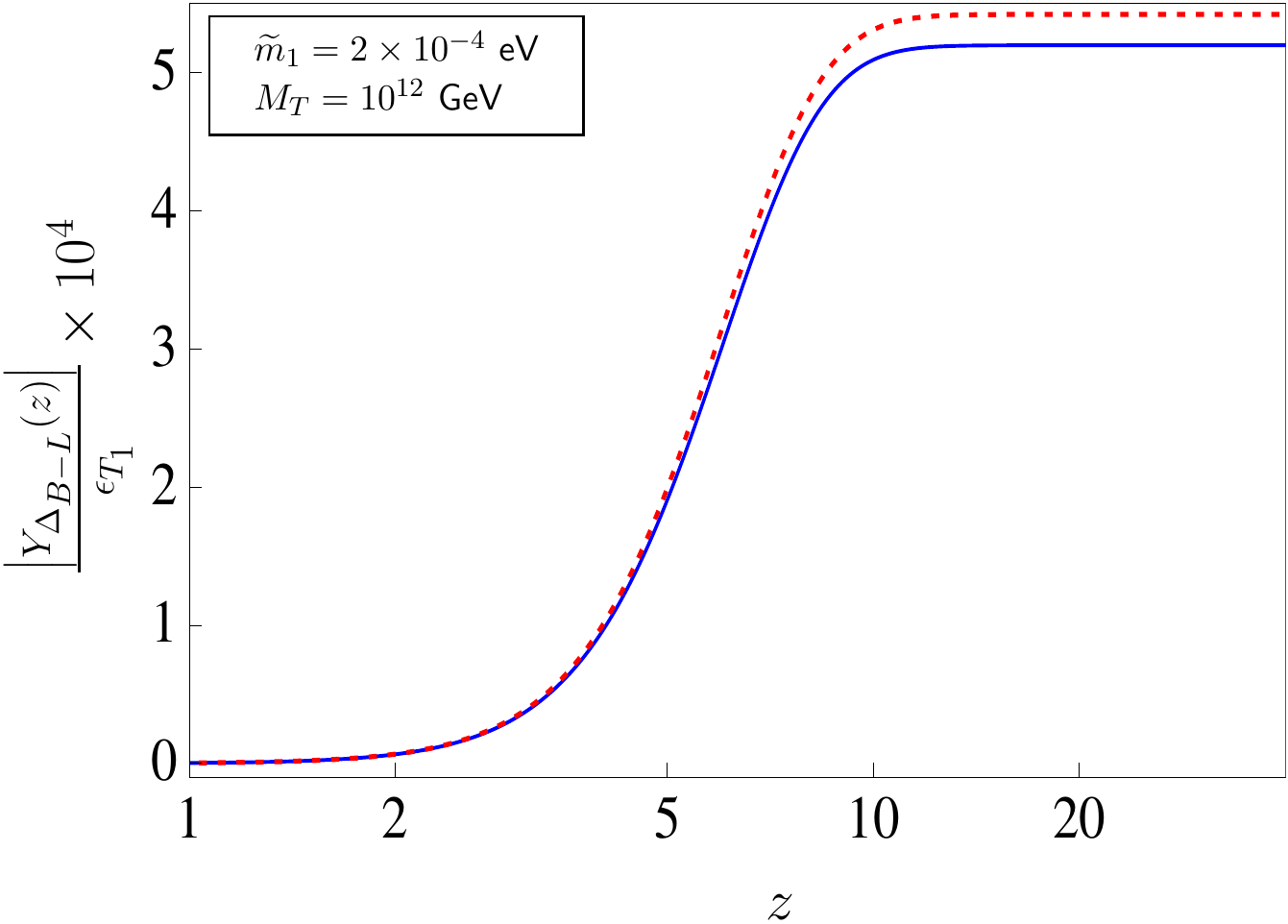}
  \includegraphics[width=7.8cm,height=6cm]{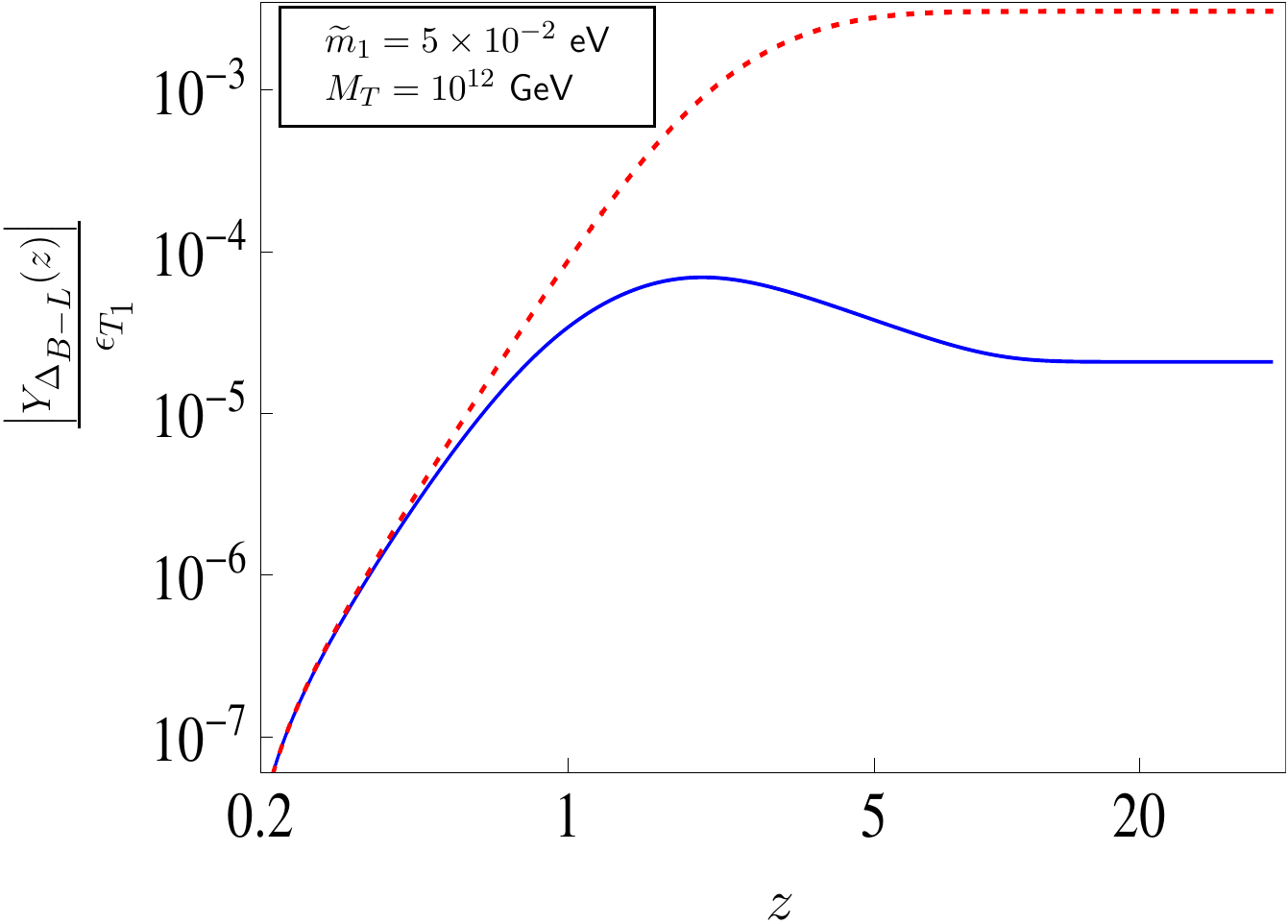}
  \caption{$Y_{\Delta_{B-L}}(z)/\epsilon_{T_\alpha}$ vs $z$ for
    aligned (solid blue line) and flavored (dotted red line) cases in
    the temperature regime $10^{10}\;\mbox{GeV}\;\lesssim T\lesssim
    10^{12}\;\mbox{GeV}$.  The flavored CP asymmetries are fixed as
    $\epsilon_{T_1}^{\ell_1}=-0.1\times \epsilon_{T_1}$,
    $\epsilon_{T_1}^{\ell_\tau}=1.1\times \epsilon_{T_1}$ and
    $\epsilon_{T_1}=10^{-5}$ whereas the flavor projectors are accordingly
    $K_{11}=0.99$ and $K_{1\tau}=0.01$.}
  \label{fig:asymmetries-flav-unflav}
\end{figure}
In order to show this is in fact the case, we have calculated the
evolution of the $Y_{\Delta_{B-L}}(z)$ asymmetry for $M_{T_1}=10^{12}$
GeV and $\widetilde m_1=2\times 10^{-4},5\times 10^{-2}$ eV by
numerically integrating eqs.~(\ref{eq:BEQs1}) and (\ref{eq:BEQs2}),
and assuming the following flavor configuration $K_{11}=0.99$,
($K_{\tau 1}= 1 - K_{11}$) and $\epsilon_{T_1}^{\ell_1}=-0.1\times
\epsilon_{T_1}$ and $\epsilon_{T_1}^{\ell_\tau}=1.1\times
\epsilon_{T_1}$ with $\epsilon_{T_1}=10^{-5}$. The result is displayed
in fig.~\ref{fig:asymmetries-flav-unflav} where, in addition to the
resulting asymmetry in the flavored case, we also show the
corresponding asymmetry in the fully aligned case to facilitate the
comparison between both cases. As can be seen in fig.
\ref{fig:asymmetries-flav-unflav} (left panel) for $\widetilde
m_1=2\times 10^{-4}$ eV (regime $(a)$) flavor effects are small,
producing only a $\sim 5\%$ enhancement. In contrast, for $\widetilde
m_1=5\times 10^{-2}$ eV, and given that the flavor configuration we
chose satisfies~(\ref{eq:flavor-enhancement-cond}), flavor effects
produce a two orders of magnitude enhancement of the final $B-L$
asymmetry: $Y^{\text{Alig}}_{\Delta_{B-L}}=2.1\times 10^{-10}$ whereas
$Y^{\text{Flav}}_{\Delta_{B-L}}=3.1\times 10^{-8}$. As in the standard
case the effects of flavor could be even larger once the muon Yukawa
coupling enters into thermodynamical equilibrium since in that case
two flavor projectors can be simultaneously small.

\begin{figure}
  \centering
  \includegraphics[height=6.5cm,width=9cm]{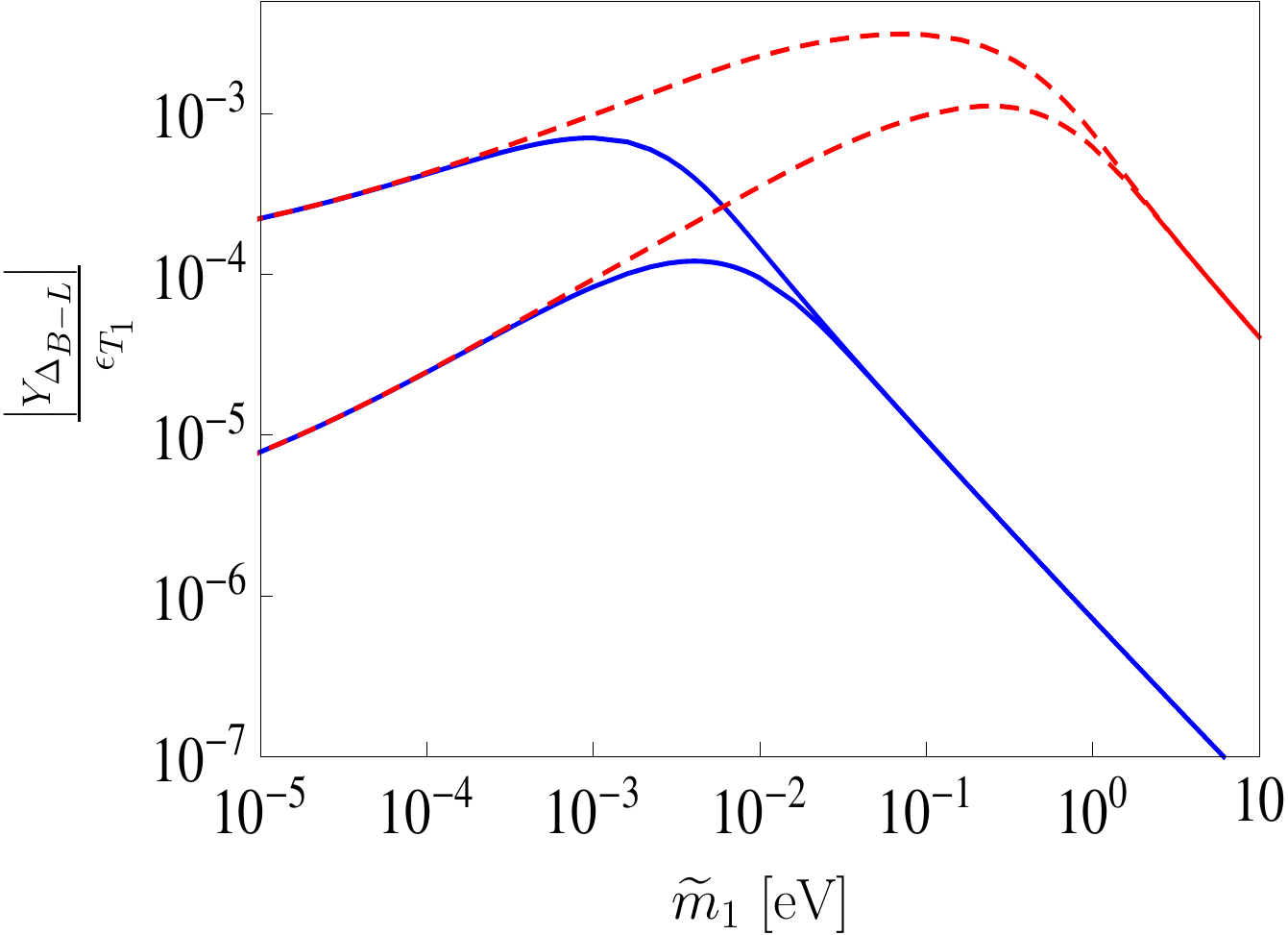}
  \caption{$Y_{\Delta_{B-L}}/\epsilon_{T_1}$ as a function of
    $\widetilde m_1$ for flavored (dotted red lines) and aligned
    (solid blue line) cases. The upper (lower) dotted and solid curves
    correspond to $M_{T_1}=10^{12}$ GeV ($M_{T_1}=10^{10}$ GeV). The
    flavored CP asymmetries and flavor projectors were fixed as in
    fig.~\ref{fig:asymmetries-flav-unflav}.}
  \label{fig:asymmetries-flav-unflav-mtilde}
\end{figure}

As previously discussed, the minimum value of $\widetilde m_1$ for
which flavor effects can produce a sizable enhancement of the $B-L$
asymmetry is not unique and depends on the triplet mass. Fig.
\ref{fig:asymmetries-flav-unflav-mtilde} shows the behavior of the
$B-L$ asymmetry with $\widetilde m_1$. As can be seen for
$M_T=10^{12}$ GeV flavor effects start playing a role at $\widetilde
m_1\sim 10^{-3}$ eV and produce a strong enhancement above this value
i.e. well inside the ``Yukawa region'' (see
fig. \ref{fig:eff-mtilde-aligned} right panel). For $M_T=10^{10}$ GeV,
as anticipated, the related effects start being relevant for larger
values of $\widetilde m_1$, again well within the ``Yukawa
region''.
%

\subsection{Implications for TeV leptogenesis}
\label{sec:implications-TeV-lepto}
TeV triplets are thermalized by gauge boson mediated annihilations up
to $z\gg 1$ (see fig. \ref{fig:guage-yukawa-reaction-dens}). The
generation of the $B-L$ asymmetry in that case proceeds basically
above this $z$ once the relic fraction that survives annihilation
start decaying. Sphaleron interactions transform this asymmetry into a
$B$ asymmetry up to temperatures $T_{\text dec}$ at which their
reactions are suddenly decoupled by the spontaneous breaking of the
$SU(2)$ symmetry~\cite{Burnier:2005hp}.  This constraint combined with
$Y_{\Delta_B}\sim 10^{-10}$ implies the bound $M_T\gtrsim$1.6 TeV
\cite{Strumia:2008cf}.  This is to be compared with the standard
resonant leptogenesis framework in which the fermionic singlet can
have a mass well below the TeV scale. The reason is that while in the
standard case the efficiency is basically determined by $\widetilde
m$, in the fermionic triplet scenario there is a dependence on $M_T$
that strongly suppress the efficiency when $M_T\sim{\cal
  O}(\mbox{TeV})$.

Adding flavor effects in the analysis does not allow to relax the
bound: in sec.~\ref{sec:flavor-effects} we argued that flavor effects
can yield a relevant enhancement of the $B-L$ asymmetry (flavored
efficiencies) only when gauge interactions decouple before Yukawa
reactions, which requires large values of $\widetilde m$, as shown in
fig.~\ref{fig:eff-mtilde-aligned} (right panel). In order to discuss
how this could take place for TeV triplets let us fix $M_T=10^3$ GeV.
Since the sphaleron decoupling temperature is determined by the Higgs
mass ($m_h$) according to $T_{\text{dec}}\simeq[80 +0.45
(m_h/\mbox{GeV})]$~GeV (with $m_h \in[114,200]$~GeV)
\cite{Strumia:2008cf,Burnier:2005hp}, fixing $m_h=120$ GeV sphaleron
decoupling will take place at $z_{\text{dec}}\sim 7.5$. At this $z$
Yukawa reactions will overcome gauge interactions provided $\widetilde
m\gtrsim 3.5$ keV \footnote{Such values are possible only if the
  different Yukawa couplings $\lambda$ are highly fine-tunned.},
however since no asymmetry could be produced until inverse Yukawa
decays decouple flavor effects are irrelevant, as this will take place
much more above $z_{\text{dec}}$.

Given that the LHC reach for fermion triplets discovery is around
700-800~GeV~\cite{Franceschini:2008pz,Arhrib:2009mz}, from the
previous discussion, one could be tempted to conclude that any
observation of fermionic triplets at the LHC would rule out
leptogenesis within this framework. This however is not entirely
correct as the asymmetry might be generated by degrees of freedom at
higher scales -- $T_2$ in triplet scenario for example\footnote{This
  possibility has been studied in the standard case in both, the one
  flavor approximation~\cite{DiBari:2005st} and flavored~\cite{Engelhard:2006yg,Antusch:2010ms} cases.}. In this case, since
$T_1$ will not play any role in the generation of the $B-L$ asymmetry,
the lower limit on $M_{T_1}$ will no longer hold. However, although
$T_1$ does not participate in leptogenesis, the asymmetry might still
be erased by $T_1$ flavor dynamics. The particular flavor structures
required to prevent too large $T_1$ related washouts may yield
particular collider signals that can be used to determine whether
leptogenesis might be relevant for the generation of the cosmic baryon
asymmetry. In the following we will discuss such a possibility.

Without the loss of generality, we assume the asymmetry is produced by
heavy fermion triplet ($T_2$) dynamics at scales $10^{9}\mbox{ GeV}
\lesssim M_{T_2} \lesssim 10^{12}\,\mbox{GeV}$. Requiring $M_{T_1}
\lesssim 1$ TeV the flavored CP violating asymmetries
$\epsilon_{T_1}^{\ell_j}$ are negligible (see eq.~(\ref{eq:bound}))
and accordingly $T_1$ decays do not generate any asymmetry. Thus,
given this triplet mass spectrum, the asymmetry is generated by $T_2$
decays (at $z_2=M_2/T\sim 1$) in the two-flavored regime $(\ell_\tau,
\ell_2)$, where $\ell_2$ is an admixture of muon and electron
flavors. The calculation of the $B-L$ asymmetry at this stage goes
along the same lines as in the case we discussed at the beginning of
this section, but taking into account that in this case we deal with
$T_2$ dynamics. At $z\ll z_2$, once muon Yukawa interactions enter in
thermodynamical equilibrium, the $Y_{\Delta_\tau}$ produced at $z \sim
z_2$ remains frozen whereas $Y_{\Delta_2}$ splits into electron and
muon asymmetries. The splitting, in the case of
$Y_{T_2}^{\text{in}}=0$, is determined by~\cite{Antusch:2010ms}
\begin{equation}
  \label{eq:splitting}
  Y_{\Delta_k}=\frac{K_{k2}}{\widetilde{K}_{22}}Y_{\Delta_2}\,,
\end{equation}
with $k=e,\mu$ and $\sum_{k=e,\mu} K_{k2}/\widetilde{K}_{22}=1$. 

As the temperature drops and reaches $z_1=M_{T_1}/T\sim 1$, $T_1$ related
washout effects become effective and are determined by
\begin{equation}
  \label{eq:T1-related-washout-effects}
  \frac{d Y_{\Delta_i}(z_1)}{dz_1}=-\frac{K_{i1}}{2\,s\,H\,z_1\,Y_\ell^{\text{Eq}}}
  \sum_{j=e,\mu,\tau}C_{ij}^\ell Y_{\Delta_j}(z_1)\gamma_{D_1}
  =-\frac{\kappa_{i1}}{4}\sum_{j=e,\mu,\tau} C^\ell_{ij}
  Y_{\Delta_j}(z_1)K_1(z_1)z_1^3\,,
\end{equation}
where $K_1(z_1)$ is the modified Bessel function of first-type, the
parameter $\kappa_{i1}=\widetilde m_{i1}/m_\star$, with $m_\star=8\pi
v^2 H|_{z_1=1}/M_{T_1}^2 = 2.25 \times 10^{-3}$~eV, and the flavor coupling matrix at this
stage is given by~\cite{Nardi:2006fx}
\begin{equation}
  \label{eq:coupling-matrix-T2lepto}
  C^\ell=\frac{1}{711}
  \begin{pmatrix}
    221 & -16 & -16\\
    -16 & 221 & -16\\
    -16 & -16 & 221
  \end{pmatrix}\,.
\end{equation}
A rough estimate of washout effects can be done by simply neglecting
flavor coupling i.e. setting $C^\ell=\bf I$ and assuming the $T_1$
related washout processes are efficient before sphalerons decouple
below the electroweak phase transition. In that case
eq. (\ref{eq:T1-related-washout-effects}) can be analytically
integrated and the final baryon asymmetry can be written in terms of
flavored asymmetries ($Y_{\Delta_i}^{\text{in}}$) generated at a high
scale (in $T_2$ decays) and $T_1$ washout related damping factors
\begin{equation} \label{eq:YinWashout}
  Y_{\Delta_B} \simeq  \sum_{i=e,\mu,\tau} Y^{\text{in}}_{\Delta_i}
  e^{-3\pi\kappa_{i1}/8} \,.
\end{equation}
Whether the correct amount of baryon
asymmetry can be generated therefore depends not only on
$Y_{\Delta_i}^{\text{in}}$ but also on the $T_1$ related washout parameters
$\kappa_{i1}$. In particular, for any value of  $Y_{\Delta_i}^{\text{in}}$, there exist $\kappa_{i1}>\kappa^{\text{max}}_{i1}$,  for which that particular flavored asymmetry will be completely washed out by $T_1$ dynamics before sphaleron decoupling and will thus not contribute to $Y_{\Delta_B}$. Including flavor dynamics and finite sphaleron decoupling temperature, the values of $\kappa_{i1}^{\text{max}}$ must be determined numerically, by solving eqs. (\ref{eq:T1-related-washout-effects}) for a given set of $Y_{\Delta_i}^{\text{in}}$ and $M_{T_1}$. Generically however, taking into account
that $T_1$ related washouts are relevant only if $\kappa_{i1}\gtrsim 1$ we can already
distinguish between three cases:
\begin{enumerate}
\item\label{case1} $\pmb{\kappa_{i1}\ll 1}$ {\bf for all flavors}. In that case
 all $T_1$ washout processes are weak and the baryon asymmetry is
  determined by high scale dynamics (of $T_2$ in our scenario). 
  Successful leptogenesis is possible in principle depending on the (practically unmeasurable) high-scale ($T_2$ related) parameters.
\item\label{case2} $\pmb{\kappa_{i1}\ll 1}$ {\bf and}
  $\pmb{\kappa_{j1}\gtrsim 1}$.  The asymmetry in flavor(s) $j$ will now be
  subject to possibly strong washouts due to $T_1$. However, the asymmetry in flavor(s) $i$ will again be
  completely determined at the high scale (by $T_2$ decays).
 In general therefore any
  constraints on $\kappa_{j1}$ rely on the assumptions about the size of
  $Y_{\Delta_i}^{\text{in}}$. Even if some of $Y_{\Delta_j}^{\text{in}}$ would be
  strongly damped, leptogenesis might (or not) account
  for the baryon asymmetry.
\item\label{case3} $\pmb{\kappa_{i1}\gtrsim 1}$ {\bf for all
    flavors}. Washout effects are relevant in all the flavors and
 the relevant washout parameters can in principle be
  constrained, depending on the size of 
  $Y_{\Delta_i}^{\text{in}}$. 
  Once for a given flavor $\kappa_{i1}$
  reaches $\kappa_{i1}^{\text{max}}$ the asymmetry in such a flavor
  will not contribute to $Y_{\Delta_B}$ implying that the final baryon
  asymmetry can be given by
  \begin{enumerate}
  \item\label{singleF} Single flavor:
    $\kappa_{i1}<\kappa_{i1}^{\text{max}}$ and
    $\kappa_{(j,k)1}>\kappa_{(j,k)1}^{\text{max}}$
  \item\label{twoF} Two flavors:
    $\kappa_{(i,j)1}<\kappa_{(i,j)1}^{\text{max}}$ and
    $\kappa_{k1}>\kappa_{k1}^{\text{max}}$
  \item\label{threeF} Three flavors:
    $\kappa_{(e,\mu,\tau)1}<\kappa_{(e,\mu,\tau)1}^{\text{max}}$
  \end{enumerate}
\end{enumerate}

\begin{figure}[!t]
\begin{center}
\includegraphics[width=9cm]{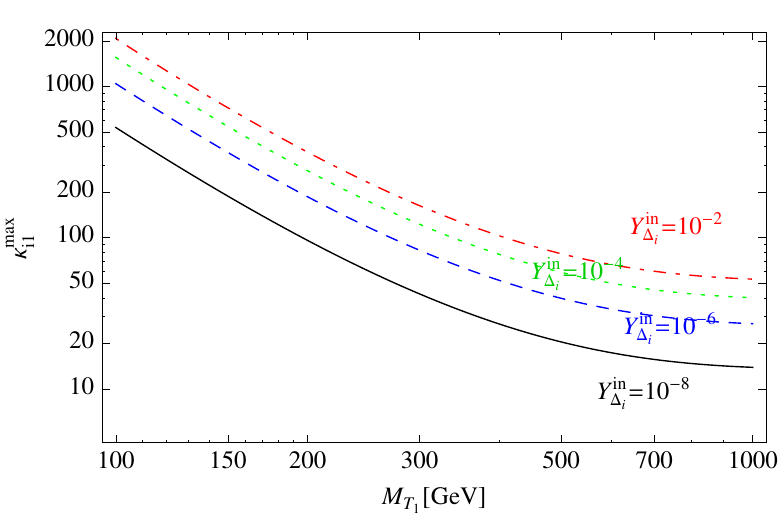}
\end{center}
\vspace{-0.7cm}
\caption{\label{fig:mTkmax} Solutions of $\kappa_{i1}^{\text{max}}$ for
  given values of high-scale generated lepton asymmetries
  $Y^{\text{in}}_{\Delta_i}$. See text for details.}
\end{figure}

Finally, in order to determine $\kappa_{i1}^{\text{max}}$ we solve
numerically eqs. (\ref{eq:T1-related-washout-effects}) for given
individual $Y_{\Delta_i}^{\text{in}}$ ranging from zero up to
$10^{-2}$ (the largest values are motivated by completeness in order
to cover possible extreme cases such as resonant leptogenesis and
possibly large flavor enhancements) and require the final asymmetry
$Y_{\Delta_B}(T_{dec.})$ to satisfy the experimental constraint
$Y_{\Delta_B}=[8.52,8.98]\times 10^{-11}$. We plot the obtained
solutions (which are independent of lepton flavor, as can be inferred
directly from the lepton flavor permutation symmetry of
eq.~(\ref{eq:T1-related-washout-effects})) as contours in the plane of
$M_{T_1}$ and $\kappa_{i1}^{\text{max}}$ in
figure~\ref{fig:mTkmax}.  We see that successful high scale thermal
leptogenesis ($Y^{\text{in}}_{\Delta_i}\lesssim 10^{-8}$) requires
$\kappa_{i1}<10(500)$ for the $T_1$ masses of
$M_{T_1}=1000(100)$~GeV. On the other hand, in the extreme case
$Y^{\text{in}}_{\Delta_i}\sim 10^{-2}$, $T_1$ washout parameters as large as
$\kappa_{1i}\lesssim 50(2000)$ are allowed for $T_1$ masses of
$M_{T_1}=1000(100)$~GeV respectively.

\subsection{Implications for type III see-saw models at Colliders}
\label{sec:3.2}

Fermionic triplets responsible for neutrino mass generation could be produced
directly at colliders and thus shed light on the viability of high scale leptogenesis
in type III seesaw scenarios. Existing studies of triplet production at the LHC
~\cite{Franceschini:2008pz,delAguila:2008cj,Arhrib:2009mz} discuss the signals and
backgrounds in detail. The main conclusion is that a fermionic triplet
with a mass up to 700 GeV would be accessible at the LHC running at $\sqrt s = 14 \text{ TeV}$
with an integrated luminosity of $100 \text{ fb}^{-1}$. The 'smoking gun' signal would be the lepton
number violating channel of two same sign leptons and four jets without missing energy.
The main SM background comes from $t \overline t$ plus jet events which
can be sufficiently suppressed by appropriate cuts on $p_T$ and missing energy~\cite{Franceschini:2008pz,Arhrib:2009mz}.

The flavored decay rates of the triplet are proportional to its Yukawa couplings to light leptons. 
Therefore, small Yukawa values imply a long lived particle, resulting in displaced 
secondary vertices in collider detectors. In type III scenarios, the proper decay length of the lightest triplet can range up to $\sim10\text{ mm}$ in the minimal setups~\cite{Arhrib:2009mz} and more than a meter if three 
triplets are introduced alltogether~\cite{Franceschini:2008pz}.

\begin{figure}[!t]
\begin{center}
\includegraphics[width=9cm]{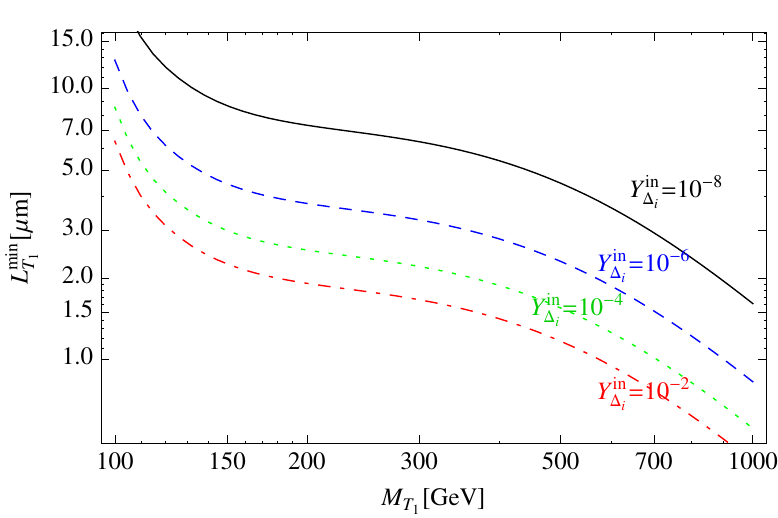}
\end{center}
\vspace{-0.7cm}
\caption{\label{fig:mTkmax1} Minimum total $T_1$ decay length versus the $T_1$ mass for
  given values of high-scale generated lepton asymmetries
  $Y^{\text{in}}_{\Delta_i}$. See text for details.}
\end{figure}

Obviously in the limit of vanishing Yukawa couplings, also the associated washout effects vanish as seen in 
eq.~\eqref{eq:YinWashout}. In order to quantify the threshold, where the
washout becomes relevant we use eq.~(\ref{eq:decay-width}) for 
the triplet partial decay width to a given lepton flavor. The expression is written in an $SU(2)$ invariant way and thus valid in the heavy triplet limit ($M_{T_1}\gg v$). Close to the electroweak symmetry breaking scale, one needs to work in the mass eigenbasis and take into account kinematic corrections due to finite $W, Z$ and Higgs masses resulting in~\cite{Bajc:2006ib}
\begin{eqnarray} \label{eq:partial-decay-width}
\Gamma(\ell_i) &\simeq& 10^{-12}\;
\kappa_i\; \left( \frac{M_{T_1}}{1~\mathrm{TeV}} \right)^2 \frac{1}{4} \left[ 2 f_1 \left(\frac{m_W}{M_{T_1}}\right) + f_1 \left(\frac{m_Z}{M_{T_1}}\right) + f_0 \left(\frac{m_h}{M_{T_1}}\right) \right]
\mbox{GeV}\,,\\
f_n(x) &=& (1-x^2)^2 (1+ 2 n x^2)\Theta(1-x)\,.
\end{eqnarray}
Inserting $\kappa_{i1}^{\max}$ derived in the previous section into the above triplet
decay rate (and summing over lepton flavors) gives a lower bound on the proper lifetime of a democratically decaying $T_1$ ($\kappa_e \simeq \kappa_\mu \simeq \kappa_\tau$) as a function of $M_{T_1}$ which is shown in figure~\ref{fig:mTkmax1}. We observe that kinematic effects are important for triplet masses below $150\,\mathrm{GeV}$ where also our estimates for the washout cease to be reliable. 

Although the experimental resolution of the secondary vertex in the LHC experiments strongly depends on the flavor composition of the decay, it should allow to reconstruct a secondary vertex due to a triplet to no better than around $100 \,\mu\text{m}$ in the direction orthogonal to the beam axis~\cite{Franceschini:2008pz}. Therefore, seeing a triplet induced displaced vertex at the LHC certainly implies cases \ref{case1} or 3(c) of the previous section -- {\it {a long lived triplet accessible at the LHC is consistent with high scale leptogenesis accounting for the cosmic baryon asymmetry.}}

On the other hand, completely excluding a pre-existing leptonic asymmetry at the LHC seems quite
challenging, even for $Y_{\Delta_i}^{\text{in}} = 10^{-8}$ as allowed by high scale
thermal non-resonant leptogenesis. Cases \ref{case2}, \ref{singleF} and \ref{twoF} 
are all characterized by a relatively short lived triplet with its proper length below $10\mu \text{m}$ 
for $M_{T_1} > 150 \text{ GeV}$. The maximum proper length excluding the survival of any lepton asymmetries for a given triplet mass is reached for approximately equal decay branching ratios into all lepton flavors. The required improvement in the secondary vertex resolution for this case might be offered by the proposed ILC experiment \cite{Behnke:2007gj}. Triplets of a couple hundred GeV may be produced copiously at the ILC with a total cross section $\sigma \sim \mathcal O(100 \text{ fb})$~\cite{Arhrib:2009mz}. If the proposed detector requirements~\cite{Behnke:2007gj} were realized, the vertexing resolution would be roughly $1/3$ of the LHC's detectors which is close to the interesting region discussed above\footnote{Note that in the detector reference frame, decay lengths
of particles of a couple of hundred GeV may receive sizable boost factors.}~\cite{Behnke:2007gj}. 

\begin{figure}[t]
\begin{center}
\includegraphics[width=9cm]{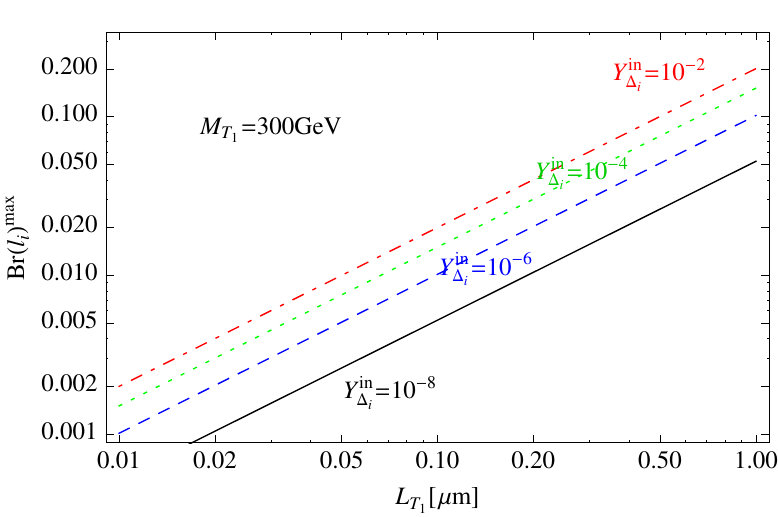}
\end{center}
\vspace{-0.7cm}
\caption{Parameter space regions of $T_1$, allowing for successful
  high scale leptogenesis yielding the given asymmetry values in
  lepton flavor $i$. Regions below the curves are allowed for each
  $Y^{\text{in}}_{\Delta_i}$ value.  $Br(\ell_i
  )^{\mathrm{max}}$ stands for the maximum allowed branching ratio of
  $T_1$ to a lepton of flavor $i$. See text for details.}
\label{fig:BrmT}
\end{figure}

Finally, if one or more of the flavored branching ratios is suppressed, a lepton asymmetry may survive in that particular channel. In this case, the $T_1$ lifetime may be even smaller than indicated in fig.~\ref{fig:mTkmax1}, beyond any realistic detector resolution. The condition of surviving lepton asymmetry can then be phrased in terms of an upper bound on the smallest flavored branching ratio, depending on the $T_1$ mass and total width. Experimentally the least favorable case is in the transition region between the two regimes, where the proper decay length of the triplet is below the experimental resolution
while the hierarchy between the different flavored branching ratios is
not very pronounced, making definite statements about the viability of high
scale leptogenesis difficult.  This situation is visualized in fig.~\ref{fig:BrmT} for representative
values of $Y^{\text{in}}_{\Delta_i}$ and for $M_{T_1}=300$~GeV.
Note also that in order to discriminate between the cases
\ref{singleF} and \ref{twoF}, and the parameter region where high
scale leptogenesis is ruled out, all three leptonic decay channels of
$T_1$ would need to be measured or at least experimentally
constrained.

\subsection{Implications for lepton flavor violation in minimal type III (+I) see-saw models}
\label{sec:3.3}

Recently, it has been shown, that a mixed type I+III see-saw model
with a single fermionic singlet and triplet below the TeV scale
naturally arises in a particular grand unified
scenario~\cite{Bajc:2006ia,Bajc:2006ib}. It belongs to a class of
so-called `minimal' type I and III see-saw scenarios, where the light
neutrino mass matrix due to the mixing with only two heavy Majorana
states is of rank 2 and the lightest neutrino remains (almost)
massless. A particularity of such models is that the neutrino mass
scale is fixed and that all the interactions of the heavy mediators
are fixed in terms of the measured neutrino oscillation data, a single
physical Majorana phase and a complex parameter
$z$~\cite{Ibarra:2003up}, which is the single complex rotation angle
in the $2\times 2$ orthogonal matrix $\pmb{R}$ introduced in
eq.~(\ref{eq:casas-ibarra}) (we employ conventions as defined
in~\cite{Kamenik:2009cb}). The presence of TeV scale triplets can
induce lepton flavor violating (LFV) processes which, in the minimal
models, are all correlated and their magnitude scales exponentially
with $\text{Im} \, z$. The greatest sensitivity is exhibited in the
$\mu - e$ sector by the nuclear conversion experiments, yielding
present bounds of the order $\text{Im} \, z < 7$ for the lightest
triplet mass of $M_T=100$~GeV~\cite{Kamenik:2009cb}.  The proposed
next generation PRISM/PRIME experiments~\cite{Ankenbrandt:2006zu} are
expected to improve the sensitivity down to $\text{Im} \, z < 4$ for
the same triplet mass.

\begin{figure}[t]
\begin{center}
\includegraphics[width=7.5cm,height=5.7cm]{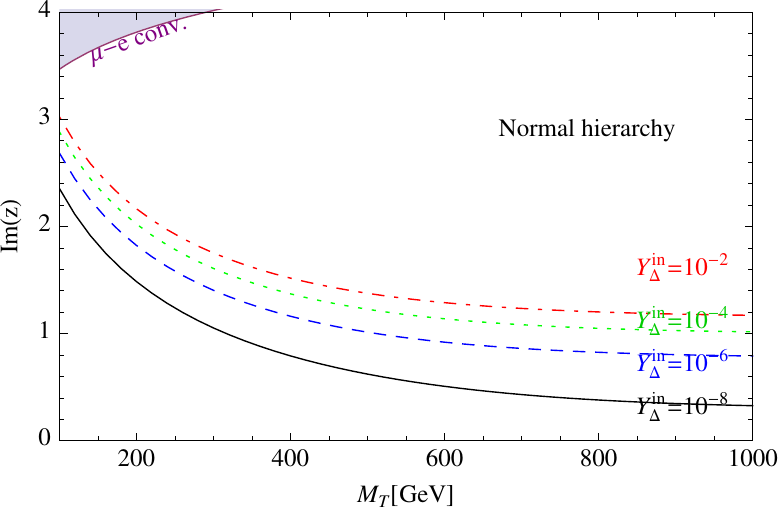}
\includegraphics[width=7.5cm,height=5.7cm]{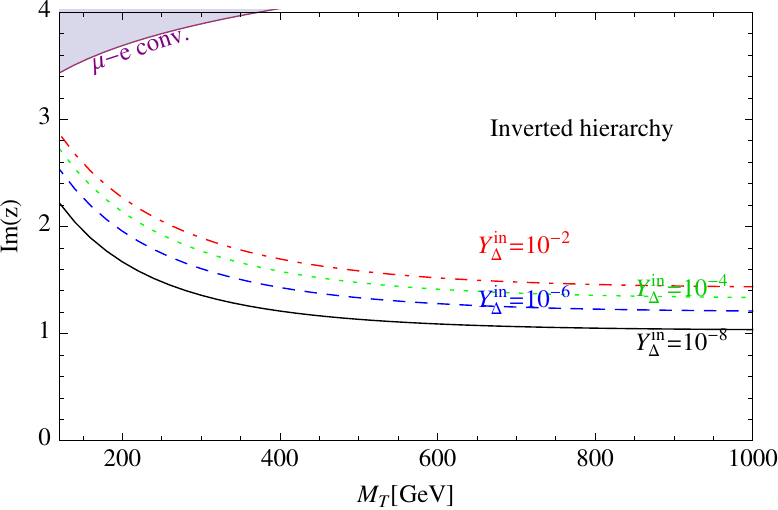}
\end{center}
\vspace{-0.7cm}
\caption{Correlation of leptogenesis bounds (area below the contours
  is allowed for a given $Y_{\Delta}^{\text{in}}$) and projected $\mu
  - e$ conversion sensitivity of PRISM/PRIME~\cite{Ankenbrandt:2006zu}
  (shaded area) on the parameters of the minimal I+III model
  ($\text{Im} z$ and $M_T$). See text for details.}
\label{fig:LFV}
\end{figure}

From the discussion in the previous sections it is clear, that minimal
type III models at the TeV scale cannot mediate successful
leptogenesis\footnote{It has been proposed that resonant leptogenesis
  could be viable in the TeV scale minimal I+III
  model~\cite{Bajc:2006ib}. At first glance, it seems like an
  interesting possibility, since the fermionic singlet mediator would
  not suffer from the gauge reaction related strong washouts plaguing
  TeV triplet leptogenesis. However, at temperatures above the
  electroweak phase transition the mixing between $SU(2)$ singlets and
  triplets is highly suppressed and a large resonant enhancement due
  to the interference with the wavefunction diagram in figure 1 cannot
  occur~\cite{Blanchet:2008cj}.}. However, the requirements that the flavor dynamics of the
TeV scale triplets do not wash out any lepton number asymmetries
generated at a higher scale by other mechanisms can in these models be
phrased entirely in terms of $M_T$ and $z$. As such they can then be
correlated with the projected sensitivity of future LFV
experiments. In particular for the minimal I+III see-saw scenario we
plot the constraints imposed by allowing for a high scale lepton asymmetry to survive triplet related washouts
for a given $Y_{\Delta}^{\text{in}} \equiv \sum_i
Y_{\Delta_i}^{\text{in}}$ on the parameters of the model ($\text{Im}
\, z$ and $M_T$) and superimpose them onto the projected sensitivity
of future $\mu-e$ conversion experiments. In this procedure we
marginalize over unknown phases, individual $Y_{\Delta_i}^{\text{in}}$
(while keeping their sum fixed) and also $\text{Re} \, z$. The results
are shown in figure~\ref{fig:LFV} for normal (on the left side) and
inverted (on the right side) neutrino mass hierarchy.
The results for the minimal pure type III models are similar. We
observe that in the minimal models, requiring a high-scale lepton asymmetry to survive triplet related washouts rules out LFV effects which could be
detected in the foreseeable future. Conversely, observation of minimal
see-saw models at the LHC and detection of the associated LFV signals
in the next generation of $\mu - e$ nuclear conversion experiments
would disfavor leptogenesis as the mechanism for the generation of the
cosmic baryon asymmetry.

Finally, within minimal scenarios, LHC (and ILC) may yield additional information on the size of the Yukawa
couplings. Since in order to have an effective washout, $\text{Im } z \gtrsim 1$
is required as seen on fig. \ref{fig:LFV} for both normal and inverted hierarchies, predictions for correlations between different flavored branching ratios emerge, as discussed at length in~\cite{Arhrib:2009mz}.
For example, in the case of a small $\theta_{13}$ \cite{Bajc:2006ib} the inverted
hierachy gives a clear prediction
\begin{equation}
	\frac{\text{Br}(\tau)}{\text{Br}(\mu)} = \tan^2 \theta_{23},
\end{equation}
and measuring the electron channel may disprove a large value of $\text{Im } z$.
A similar statement, although less elegant, can be made in the case
of normal hierarchy, where $\text{Br}(e)$ and the total decay width
is a function of $z$ only.

\section{Conclusions}
\label{sec:conclusions}
The generation of a lepton asymmetry, via the decays of fermionic
$SU(2)$ triplets, can take place in two possible regimes: a ``gauge
regime'' in which gauge reactions decouple at lower temperatures than
Yukawa induced inverse decays. In that case the relic fraction of
triplets that survives the gauge boson mediated annihilation processes
will decay producing an amount of $B-L$ asymmetry whose value will
strongly depend on $M_T$.  Conversely, in the ``Yukawa regime'', at
low temperatures, the triplet distribution is driven to thermal
equilibrium by Yukawa inverse decay processes, even after gauge
interactions have switched off. In contrast to the ``gauge regime'',
in this case the amount of $B-L$ generated in $T$ decays is not
sensitive to $M_T$ and is fixed by $\widetilde m$. We have determined
the $\widetilde m-M_T$ regions that define both regimes. After
identifying these regions we have studied the effects of light lepton
flavors on the generation of the $B-L$ asymmetry. We have shown that
in the ``gauge regime'' the effects have a minor influence as they
barely reach fractions of a percent. Instead, in the ``Yukawa
regime'', as in the standard case, the enhancement of the final $B-L$
can be large, readily reaching the order of magnitude level.

We have shown that the requirement of successful $T_1$ leptogenesis
constrains the dynamics of TeV triplets to lie in the ``gauge
regime'', and therefore the inclusion of light lepton flavor effects
in the calculation of the $B-L$ asymmetry has basically no significant
effect. This implies that even after including flavor the bound
$M_{T_1}\gtrsim $ 1.6 TeV, found in the one flavor approximation~\cite{Strumia:2008cf}, still holds. Since fermionic triplets will be
produced at the LHC if their masses are below $\sim$ 1 TeV~\cite{Franceschini:2008pz} their observation will automatically
preclude them from explaining the Cosmic baryon asymmetry via $T_1$
leptogenesis. However, it will not necessarily exclude the case for
high scale leptogenesis, due to $T_2$ or other mechanisms. In fact, as
we have argued, the $B-L$ asymmetry can be built up at a very early
age ($T\ggg M_{T_1}$) and can possibly survive $T_1$ washout
processes. We have shown that $T_1$ collider observables could, in
principle, be used either to strengthen or rule out the case for high
scale leptogenesis\footnote{Note that a similar argument could apply
  for other mechanisms of high scale baryogenesis, since sphaleron
  processes will tend to equilibrate any $B$ asymmetry with $L$, which
  could in term be washed out by light Majorana triplet interactions.}
(there have been recent related attempts to
invalidate leptogenesis~\cite{Frere:2008ct} or high scale baryogenesis~\cite{Blanchet:2008zg} using collider observables). Finally, we have
shown how in minimal see-saw models, LFV observables can also be
employed in order to over-constrain the region of parameters of TeV
scale Majorana electroweak triplets allowing for successful high scale
leptogenesis.
\section{Acknowledgements}
We acknowledge many enlightening conversions on the subject with Enrico Nardi, Juan Racker and Ilja Dor\v sner. We are especially thankful to Enrico
Nardi for his suggestions and comments on the manuscript. D.A.S also wants
to thank the theoretical physics group at the Josef Stefan Institute
for their hospitality during the completion of this work and is supported by a Belgian FNRS
postdoctoral fellowship.  This work is supported in part by the
Slovenian Research Agency, by the European Commission RTN network,
Contract No. MRTN-CT-2006-035482 (FLAVIAnet) and by the Deutsche
Forschungsgemeinschaft via the Junior Research Group ``SUSY
Phenomenology" within the Collaborative Research Centre 676
``Particles, Strings and the Early Universe''.

\appendix

\section{Conventions and notation}
\label{sec:formulas}
In this appendix we fix our conventions and present the relevant
equations used in the calculations discussed in sections
\ref{sec:generalities} and \ref{sec:flavor-effects}. Some of the
formulae are quite standard, however, for completeness, we include
them. As has been mentioned we used Maxwell-Boltzmann distributions
for the equilibrium number densities, namely
  \begin{equation}
    \label{eq:eq-distributions}
    n^{\text{eq}}_T(z) =\frac{M^3_T}{\pi^2}\frac{K_2(z)}{z}\,,
    \quad n^{\text{eq}}_\ell(z)=\frac{2M_T^3}{\pi^2\;z^3}\,,
  \end{equation}
  where $z=M_T/T$ and $K_2(z)$ is the modified Bessel function of the
  second-type. With this approximation the energy density $\rho(z)$
  and pressure $p(z)$ becomes
\begin{equation}
  \label{eq:energy-dens}
  \rho(z)=\frac{3 M_T^4}{z^4\pi^2} g_*\,,\quad
  p(z)=\frac{M_T^4}{z^4\pi^2} g_*\,,
\end{equation}
where $g_*=\sum_{i=\text{All species}} g_i$ is the number of standard
model relativistic degrees of freedom (118 for $T\gg$ 300 GeV). Accordingly,
the expansion rate of the Universe and entropy density can be written
as
\begin{equation}
  \label{eq:HandS}
  H(z)=\sqrt{\frac{8g_*}{\pi}}\frac{M_T^2}{M_{\text{Planck}}}\frac{1}{z^2}\,,
  \quad
  s(z)=\frac{4 M_T^3}{z^3\pi^2} g_*\,.
\end{equation}
The decay and annihilation reaction densities in eqs.~(\ref{eq:BEQs1})
and (\ref{eq:BEQs2}) are given by
\begin{eqnarray}
  \label{eq:reaction-dens}
  \gamma_{D_\alpha}(z)&=&\frac{1}{8\pi^3}\frac{M_{T_\alpha}^5}{v^2}
  \frac{K_1(z)}{z}\widetilde m_\alpha\,,\\
  \label{eq:reaction-dens2}
  \gamma_{A_\alpha}(z)&=&\frac{M_{T_\alpha}^4}{64\pi^4}\;\int_4^\infty
  dx\;\frac{\sqrt{x}K_1(z\sqrt{x})\widehat\sigma_A(x)}{z}\,,
\end{eqnarray}
where $K_1(z)$ is the modified Bessel function of first-type and
$x=s/M^2_T$. The reduced cross section in
(\ref{eq:reaction-dens2}) accounts for the gauge boson mediated
s-channel processes $T_\alpha T_\alpha\leftrightarrow \ell\bar\ell$
and $T_\alpha T_\alpha\leftrightarrow q\bar q$ and the t and u-channel
triplet mediated process $T_\alpha T_\alpha\leftrightarrow A_\mu A^\mu$
($A^\mu=W_A^\mu,B^\mu$). For $\widehat \sigma_{A}(x)$ we used the
expression given in ref.~\cite{Hambye:2003rt}:
\begin{equation}
  \label{eq:red-cross-sec}
  \widehat \sigma_A(x)=
  \frac{6g^4}{\pi}\left(1+\frac{2}{x}\right)r
  +
  \frac{2g^4}{\pi}\left[
    3\left(
      1 + \frac{4}{x} - \frac{4}{x^2}
    \right)
    \log
    \left(
      \frac{1+r}{1-r}
    \right)
    -\left(
      4 + \frac{17}{x}
    \right)r
  \right]\,,
\end{equation}
with $r=\sqrt{1-4/x}$. 

Neglecting flavor coupling, i.e $C^\ell=I$ eq.~(\ref{eq:BEQs2}) can be
analytically integrated (by using its integrating factor). Assuming a
vanishing initial asymmetry ($Y_{\Delta_i}^{\text{in}}(z_0)=0$) the
result reads
\begin{equation}
  \label{eq:YDeltaB-Li-appendix}
  Y_{\Delta_i}(z)=-\epsilon_{T_\alpha}^{\ell_i}
  Y_{T_\alpha}^{\text{Eq}}(z_0)\eta_{i\alpha}(z)\,.
\end{equation}
The flavored efficiency function $\eta_{i\alpha}(z)$ determines the
evolution of the $\Delta_i$ asymmetry at any $z$. Using the
approximation $Y_{T_\alpha}(z)+Y_{T_\alpha}^{\text{Eq}}(z)\simeq
2Y_{T_\alpha}^{\text{Eq}}(z)$ it can be written as
\begin{equation}
  \label{eq:eff-func}
  \eta_{i\alpha}(z)=\frac{1}{Y_{T_\alpha}^{\text{Eq}}(z_0)}
  \int_{z_0}^{z}\;dz'\frac{\gamma_{D_\alpha}}
  {\gamma_{D_\alpha}+2\gamma_{A_\alpha}}\frac{dY_{T_\alpha}(z')}{dz'}
  e^{-\sum_{\alpha}\int_{z'}^z dz'' P_{i\alpha}(z'')}\;,
\end{equation}
where freeze-out of $\Delta_i$ takes place at $z_f$
($z_0\ll z_f$) and
\begin{equation}
  \label{eq:P_i_aplha}
  P_{i\alpha}(z)=\frac{K_{i\alpha}}{2 Y^{\text{Eq}}_\ell(z)}
  \frac{\gamma_{D_\alpha}(z)}{s(z) H(z) z}\;.
\end{equation}
For temperatures $T\gtrsim 10^{13}$ GeV, for which the flavor
composition of the lepton doublet states is irrelevant,
$K_{i\alpha}=1$.
 
The final $B-L$ asymmetry is determined according to
\begin{equation}
  \label{eq:finalBmL-app}
  Y_{\Delta_{B-L}}=\sum_{i=e, \mu, \tau} Y_{\Delta_i}=
  3\times \sum_{i=e,\mu\tau}\epsilon_{T_\alpha}^{\ell_i}
  Y_{T_\alpha}^{\text{Eq}}(z_0)\eta_{i\alpha}(z_f)\;,
\end{equation}
where $\eta_{i\alpha}(z_f)$ defines the efficiency factor for flavor $i$
and the factor 3 comes from the triplet degrees of freedom.

\end{document}